\documentclass[preprint,aps,prb]{revtex4}
\usepackage{bm}
\usepackage{epsfig}
\usepackage{graphicx}
\newcommand{\nix}[1]{}

\begin{document}

%\begin{frontmatter}

\title{Magneto-gyrotropic photogalvanic effects in GaN/AlGaN two-dimensional systems}

\author{W.~Weber, S.~Seidl, S.N.~Danilov, W.~Prettl, S.D.~Ganichev}
%%\corauth[cor]{Corresponding author. Tel.: +49 941 943-2050}
%%\ead{sergey.ganichev@physik.uni-regensburg.de}

\affiliation{Terahertz Center, University of Regensburg, 93040
Regensburg, Germany}
\author{V.V. Bel'kov, L.E.~Golub, E.L.~Ivchenko}%%
\affiliation{A.F.~Ioffe Physico-Technical Institute, Russian
Academy of Sciences, 194021 St.~Petersburg, Russia}
\author{Z.D. Kvon}%%
\affiliation{Institute of Semiconductor Physics, Russian Academy
of Sciences, 630090 Novosibirsk, Russia}
\author{Hyun-Ick Cho, Jung-Hee Lee}%%
\affiliation{Kyungpook National University, 1370 Sankyuk-Dong,
Daegu 702-701, Korea}

\begin{abstract}
{The magneto-gyrotropic photogalvanic  and spin-galvanic
effects are observed in  (0001)-oriented GaN/AlGaN heterojunctions excited by terahertz
radiation. We show that free-carrier absorption of  linearly or  circularly
polarized terahertz radiation in low-dimensional structures
causes an electric photocurrent in the presence of an in-plane
magnetic field.
Microscopic mechanisms of these photocurrents based on spin-related
phenomena are discussed.
Properties of the magneto-gyrotropic and spin-galvanic
effects specific for hexagonal heterostructures are analyzed.
}
\end{abstract}
%%\begin{keyword}
%%gallium nitride \sep two-dimensional systems \sep photogalvanic
%%effects
%% \PACS 73.21.Fg \sep 72.25.Fe \sep 78.67.De \sep 73.63.Hs
%%\end{keyword}

\maketitle
%\end{frontmatter}

\section{Introduction}

Gallium nitride is a promising semiconductor for spintronics since
long spin relaxation times are detected in this material~\cite{1}
and, if doped with manganese, it is expected to become
ferromagnetic with a Curie-temperature above room temperature~\cite{2}.
Recently it has been shown that in GaN/AlGaN low-dimensional structures a
substantial Rashba spin splitting in the electron band structure
is present, potentially allowing spin manipulation by an external
electric field~\cite{3}. The Rashba spin splitting due to
structural inversion asymmetry, which is not expected in wide-band
zinc-blende-based semiconductors, is caused in GaN wurtzite
heterostructures by a large piezoelectric effect, which yields a
strong built-in electric field at the GaN/AlGaN (0001) interface,
and a strong polarization induced doping
effect~\cite{DiCarlo,Litvinov03p155314}. Spin splitting in
$\bm k$-space ($\bm k$ is the electron wave vector) may yield
a variety of spin-dependent phenomena that can be
observed in electronic transport and optical measurements.
First indications of substantial
spin splitting came from the observation of circular photogalvanic
effect in GaN/AlGaN heterojunctions~\cite{3}.
Investigation of the circular photogalvanic effect
were extended to GaN quantum wells as well as
to GaN-based low dimensional structures under uniaxial stress
confirming the Rashba-type of spin splitting~\cite{PGECho,PGEHe}. By
weak-localization studies the magnitude of spin splitting has been
obtained showing that the splitting is comparable to that of GaAs-based
heterostructures being of the order of 0.3~meV at the Fermi
wave vector~\cite{4,5,5a}. On the other hand measurements of
Shubnikov-de~Haas oscillations revealed a substantially  larger
spin splitting of about 1~meV~\cite{SdH2}.

Here we report on the observation of the magneto-gyrotropic
photogalvanic effect (MPGE)~\cite{7,8,9} and the spin-galvanic
effect~\cite{6,10,11} in GaN/AlGaN heterostructures. Both effects have
been detected in (0001)-oriented structures in a wide range of
temperatures from technologically important room temperature to
liquid helium temperature. The microscopic origin of the spin-galvanic
effect is the
inherent asymmetry of spin-flip scattering of electrons in systems
with removed  spin degeneracy of the electronic bands in $\bm
k$-space~\cite{11}. The
magneto-gyrotropic photogalvanic effect  has  so far been
demonstrated in GaAs, InAs, and SiGe quantum wells where its
microscopic origin is the zero-bias spin separation~\cite{8,9}
which is caused by spin-dependent scattering
of electrons due to a
linear-in-$\bm k$ term in the scattering matrix elements. By the application of
an external  magnetic field, a pure spin current is converted
into an electric current.
%Besides demonstration of these two
%phenomena in GaN/AlGaN heterostructures, our measurements also give an
%evidence that the system under investigation here is of C$_{3v}$
%symmetry which is different from the symmetry of
%zinc-blende-structure based two-dimensional systems like GaAs and InAs quantum
%wells usually considered as  candidates for spintronics
%applications.

\section{Samples and experimental methods}

The experiments are carried out on GaN/Al$_{0.3}$Ga$_{0.7}$N
heterojunctions grown by MOCVD on a (0001)-oriented sapphire
substrate
%The thickness of the AlGaN layers was varied between 30~nm and 100~nm.
%An undoped 33~nm thick GaN buffer layer, which was grown under a pressure of 40~Pa at 550\r{ }C, is followed by
%an undoped GaN layer ($\sim $ 2.5~$\mu $m), grown under a pressure of 40~Pa at 1025\r{ }C. The undoped Al$_{0.3}$Ga$_{0.7}$N barrier
%was grown at 6.7~Pa and a temperature of 1035\r{ }C.
%
with the electron
mobility in the two-dimensional electron gas (2DEG) $\mu \approx 1200$~cm$^{2}$/Vs at
electron density $n_{s} \approx 10^{13}$~cm$^{-2}$
at room temperature~\cite{3}.
A high power THz molecular laser, optically pumped by a
TEA-CO$_2$ laser~\cite{6}, has been used to deliver
100~ns pulses of linearly polarized radiation with a  power of
about 10~kW at  wavelengths of $\lambda =90.5$~$\mu$m, 148~$\mu$m and 280~$\mu$m . The
radiation causes indirect optical transitions within the lowest
size-quantized subband.
The samples are irradiated along the
growth direction. Magnetic field induced photocurrents were
investigated applying both linearly  and  circularly polarized
radiation.
In experiment
the plane of polarization of linearly  polarized light
is rotated applying a crystal quartz $\lambda/2$ plate.
The circular polarization is obtained by means of
a  $\lambda $/4 plate. In this case the helicity  of the incident light $P_{circ}=
\sin 2 \varphi$ can be varied from $-1$ (left handed circular, $\sigma_-$) to $+1$ (right
handed circular, $\sigma_+)$ changing the angle $\varphi$ between
the initial polarization plane and the optical axis of the
$\lambda $/4 plate. An external magnetic field $B$  up to $1$~T is
applied parallel to the heterojunction interface. To measure the
photocurrent two pairs of ohmic contacts have been centered along
opposite sample edges (see insets in Fig.~1).  The current $J$,
generated by the light in the unbiased devices, is measured
via the voltage drop across a 50~$\Omega$ load resistor in a
closed-circuit configuration. The voltage is recorded with a
storage oscilloscope.

\section{Experimental results}

\begin{figure}[width=8cm]
\centerline{\epsfxsize 120mm \epsfbox{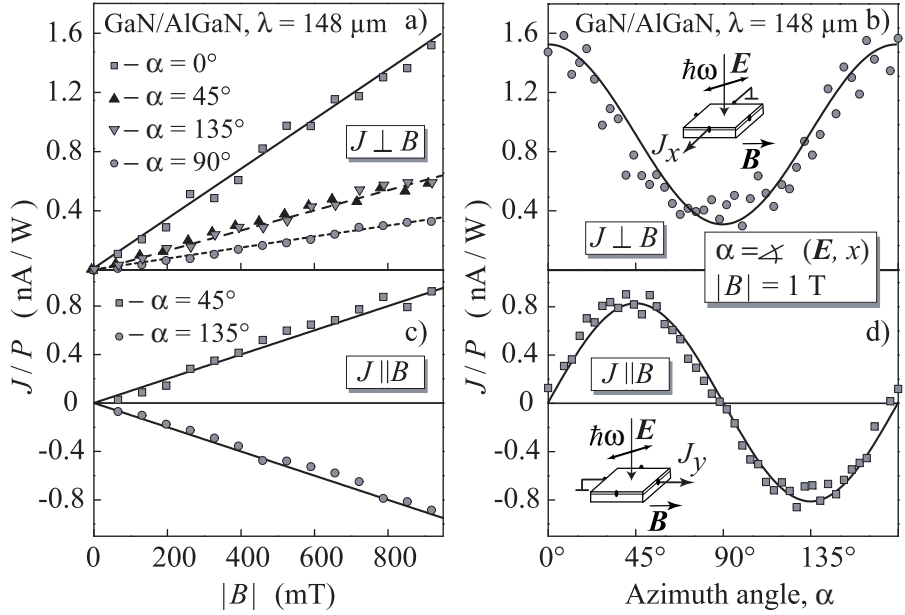}}
\caption{Magnetic field and polarization dependences of the
photocurrent $J = [J(\bm{B}) - J(-\bm{B})]/2$ measured at room
temperature in the transversal  and longitudinal geometries. a)
and b) magnetic field dependences of  the magnetic field induced
photocurrent at different azimuth angles $\alpha$, c) and d)
polarization dependences for transversal and longitudinal
geometries, respectively. Photocurrent is excited by the normally
incident linearly polarized radiation with power $P \approx
10$~kW.  Full lines are the fits  after
Eqs.~\protect(\ref{j_perp}). Insets show the experimental
geometries.} \label{fig1}
\end{figure}

Irradiating  a (0001) GaN/AlGaN heterojunction by linearly
polarized light at normal  incidence, as sketched in insets of
Fig. 1, we observe a photocurrent signal  in both perpendicular
(transversal geometry, Fig. 1a,b) and parallel (longitudinal
geometry, Fig. 1c,d) to the magnetic field $\bm B$ directions. The
width  of the current pulses is about 100~ns  which corresponds to
the terahertz laser pulses duration. The photocurrent is
proportional to the magnetic field strength and its sign depends
on the magnetic field direction. Figures~1a and~1c show the
magnetic field dependence of the photocurrent for both geometries
and for various orientation of polarization plane of linearly
polarized radiation in respect to the magnetic field direction. In
these figures
$$J = [J(\bm{B}) - J(-\bm{B})]/2$$
is plotted in order to extract the magnetic field
induced photocurrent from the total current which contains a small
magnetic field independent contribution   caused by the magnetic field independent linear
photogalvanic effect~\cite{6,10}.
Figure 1a and 1c demonstrate  that the photocurrent
exhibits an essentially  different polarization
dependence for longitudinal and transversal geometries.
While upon the variation of the azimuth angle $\alpha$
the sign of the transversal photocurrent remains unchanged, the
longitudinal current changes its direction by switching $\alpha$
from $+ 45^\circ$ to $- 45^\circ$ at constant magnetic field.
Figures~1c and~1d demonstrate the polarization dependence of the
magnetic field induced photocurrent obtained at $B = 1$~T for both geometries.
We find that the polarization dependence of the current $J$ in
transversal geometry $\bm{J} \perp \bm{B}$ is well fitted by
$J_\perp = J_1 \cos 2 \alpha + J_2$, and for the longitudinal
geometry $\bm{J} \parallel \bm{B}$ by $J_\parallel = J_3 \sin 2
\alpha $ and $J_1  \approx J_3$. Below we demonstrate that exactly
these dependences come out from the phenomenological theory.
Using two fixed polarization states in the transversal
geometry, $\alpha = 0^\circ$ and $\alpha = 90^\circ$, allows us to
extract $J_1$ and $J_2$. Adding and subtracting the currents of
both polarizations the polarization-dependent contribution $J_1$
and polarization-independent contribution $J_2$ can be obtained by
\begin{eqnarray}\label{J1J2}
  J_1 &=& \frac{J_\perp(\alpha = 0^\circ)-J_\perp(\alpha =
90^\circ)}{2}\,\, ,  \\ \nonumber
  J_2 &=& \frac{J_\perp(\alpha = 0^\circ)+J_\perp(\alpha
= 90^\circ)}{2}\,\,.
\end{eqnarray}
Figure~2 shows the temperature dependences of $J_1$ and $J_2$
together with the electron density  and mobility. It is seen that
the qualitative behaviour of both contributions is similar: at low
temperatures  they are almost independent of temperature, but at
high temperatures (for $J_1$ at $T >  100$~K and for $J_2$
at $T >  50$~K) the current strength decreases with
 temperature increasing.

\begin{figure}[width=8cm]
\centerline{\epsfxsize 90mm \epsfbox{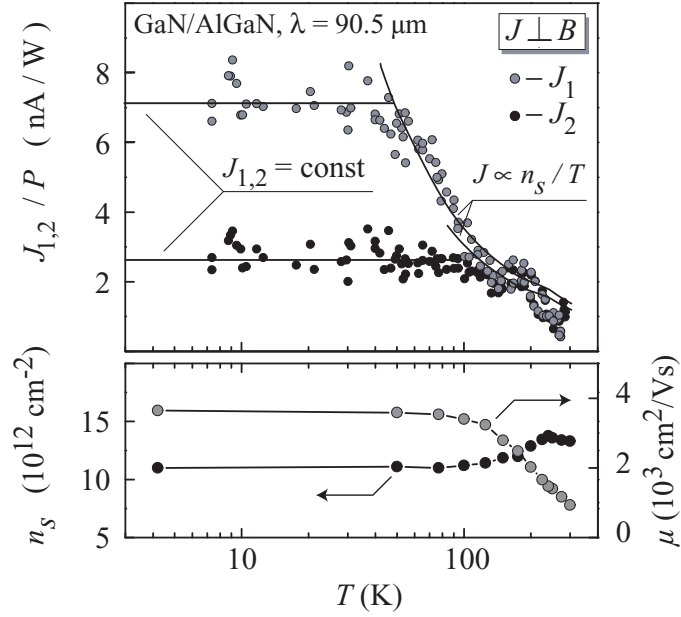}} \caption{Upper
panel: Temperature dependences  of the transversal magnetic field
induced photocurrent. Data are obtained for  $|B_y|= 0.6$~T and an
excitation at $P \approx 10$~kW. Photocurrents $J_1(T)$  and
$J_2(T)$ are obtained by subtracting and adding the currents for
the two polarizations: $J_\perp(\alpha=0^\circ)$ and
$J_\perp(\alpha=90^\circ)$ (see Eq.~(\ref{J1J2})). Full lines are
fits of $J_1(T)$ and $J_2(T)$ to $A \cdot n_s(T)/k_B T$, with a
scaling parameters $A$, and to a constant, respectively. The lower
panel shows the temperature dependences of the carrier density
$n_s$ and the electron mobility. } \label{fig3}
\end{figure}

In addition to the magnetic field induced photocurrent excited by linearly polarized radiation we also
observed a signal in response to circularly polarized light. While in the transversal geometry the
signal is insensitive to the switching of the radiation helicity from $\sigma_+$ to $\sigma_-$
in the longitudinal geometry this operation results in the reversing of the current direction. Figure 3
shows the  dependence of the photocurrent on the angle $\varphi$. These data can be well fitted by
$J_\parallel = J_4 \sin 4 \varphi + J_5 \sin 2 \varphi$. The fit, also shown in Fig. 3,  yields
the ratio $J_4 / J_5 \approx 5$ at room temperature.

\begin{figure}[width=8cm]
\centerline{\epsfxsize 75mm \epsfbox{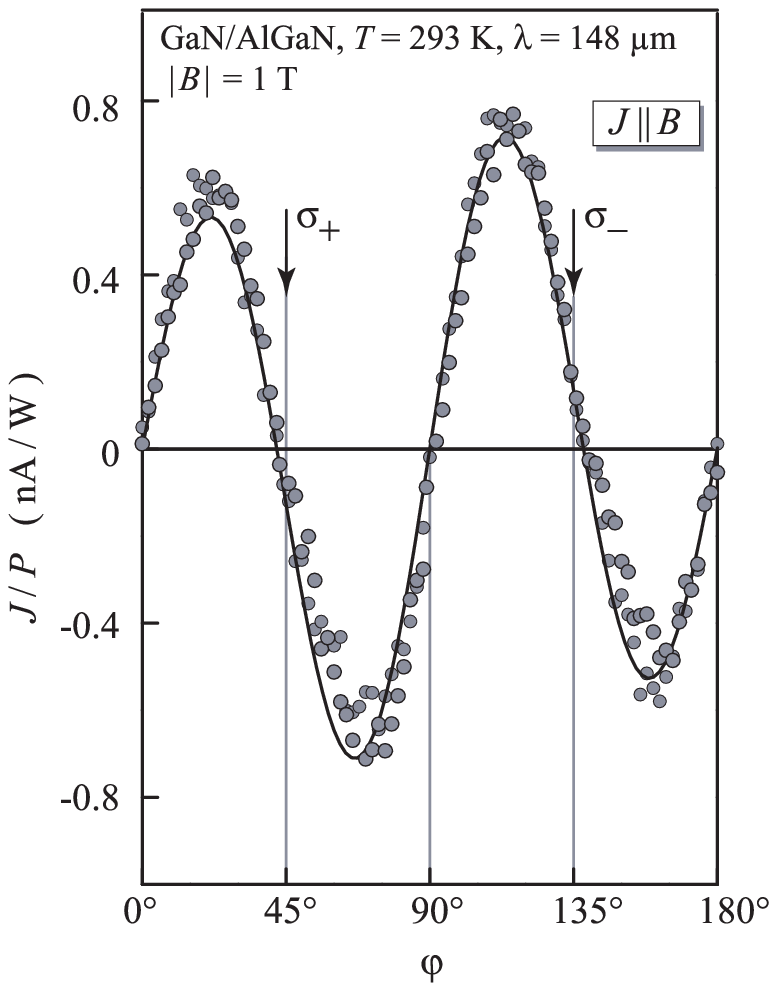}}
\caption{Magnetic field induced photocurrent $J$ as a function of
the phase angle $\varphi$ defining the radiation helicity. The
photocurrent signal is measured  at room temperature in the
longitudinal geometry $\bm J \parallel \bm B$ under normal
incidence of the radiation with $P \approx 10$~kW. The full line
is the fit after the second equation of
Eqs.~\protect(\ref{j_parallel}).} \label{fig2}
\end{figure}

\section{Discussion}

Polarization and magnetic field dependences as well as the
the difference in the photocurrent behaviour for longitudinal and transversal geometries
can be obtained in the framework of the phenomenological theory of the MPGE
which yields~\cite{7}
\begin{eqnarray}\label{definition}
\nonumber  j_\alpha &=& I \sum\limits_{\beta \gamma \delta } \phi
_{\alpha \beta \gamma \delta } B_\beta \, {e_\gamma e_\delta ^\ast
+ e_\delta e_\gamma ^\ast
\over 2} + \\
 &+& I \sum\limits_{\beta \gamma } {\mu _{\alpha \beta \gamma }
B_\beta \hat {e}_\gamma P_{circ} }\: .
\end{eqnarray}
%\begin{equation} \end{equation}
%
Here $\bm j$ is the photocurrent density proportional to $J$, $I$ the light intensity inside the heterostructure,
$\hat{\bm \phi}$ the fourth-rank tensor symmetric in the last two
indices, $\hat{\bm \mu}$ the third-rank tensor, ${\bm e}$
the light polarization  vector, and $\hat{\bm e}$  a unit vector in the
light propagation direction.
While the second term in the right hand side of the Eq.~(\ref{definition})
requires circularly polarized radiation the first term may be
non-zero for any, even unpolarized, radiation.

The point-group symmetry of (0001)-grown GaN-based low-domensional structures is
$C_{3v}$. The symmetry analysis for the $C_{3v}$ point group shows
that Eq.~(\ref{definition}) for normal light incidence reduces to
\begin{eqnarray} \label{j_xy}%
\nonumber  j_x =&& I S_1 B_y - I S_2 B_x \,\left( {e_x e_y^\ast
+e_y e_x^\ast } \right)+ \\ \nonumber &+& I S_2 B_y \,\left(
{\left| {e_x } \right|^2-\left| {e_y } \right|^2} \right)+ I \mu
B_x P_{circ}\:
,\\
 j_y =& -& I S_1 B_x +I S_2 B_y \,\left( {e_x e_y^\ast
+e_y e_x^\ast } \right)+ \\ \nonumber &+&I S_2 B_x \,\left(
{\left| {e_x } \right|^2-\left| {e_y } \right|^2} \right)+ I \mu
B_y P_{circ}\: .
\end{eqnarray}
Here three linearly independent constants are introduced as
follows: $S_{1,2} =(\phi _{xyxx} \pm \phi _{xyyy})/2$,  and
$\mu =\mu_{xxz}$.
Note that Eqs.~(\ref{j_xy}) are also valid for the $C_{6v}$ symmetry
of bulk hexagonal GaN and even for uniaxial systems of the $C_{\infty v}$
symmetry. The form of these equations is invariant under a
transfer from a given coordinate frame $(x, y)$ to another one
obtained by the rotation around the $z$ axis by any azimuth angle.
The fact that the photocurrents in both directions are described by
only three coefficients  ($S_1$, $S_2$, and $\mu$) is in contrast
to the MPGE in systems of the $C_{2v}$ symmetry~\cite{7}, where in the similar
equations all coefficients in the eight terms in the right hand sides of
Eqs.~(\ref{j_xy}) are linearly independent.

Equations~(\ref{j_xy}) can be directly applied to describe
polarization dependences in longitudinal and transversal
geometries. For the geometry used in our experiment where the
counterclockwise azimuth angle $\alpha$ between
the $x$ axis and the unit vector of linear polarization
is varied, the polarization-dependent factors in
Eqs.~(\ref{j_xy}) can be written as
$\left| {e_x } \right|^2-\left| {e_y } \right|^2=
\cos 2\alpha$,  $e_x e_y^\ast +e_y e_x^\ast = \sin 2\alpha$
and $P_{circ} = 0$.
According to Eqs.~(\ref{j_xy}) for a fixed direction of the
magnetic field, say $\bm{B} \parallel y$, the photocurrent in
response to the linearly polarized radiation is given by
\begin{eqnarray}
\label{j_perp}
j_x &=& I B_y (S_1 + S_2 \cos{ 2 \alpha}) \: ,\\
j_y &=& I B_y S_2  \sin{ 2 \alpha}\:.  \nonumber
\end{eqnarray}
Fits of experimental data to these equations are shown in Figs.~1c
and~1d demonstrating a good agreement with the theory.  We
emphasize i) the presence of a substantial contribution of the
polarization-independent first term on the right hand side of the
Eq.~(\ref{j_perp}) to the current $j_x$ and ii)  the equal amplitudes of
the $\alpha$-dependent contributions for both transversal and
longitudinal geometries.

For elliptically polarized light obtained in experiments by
varying the angle $\varphi$ between the initial polarization plane
and the optical axis of the $\lambda $/4 plate the polarization dependent terms are described by:
$\left| {e_x } \right|^2-\left| {e_y } \right|^2= (1+\cos{4\varphi})/2$, $e_x
e_y^\ast +e_y e_x^\ast = {\sin 4\varphi}/2$ and $P_{circ} = \sin
2\varphi$. Thus  for magnetic field applied along $y$,
Eqs.~(\ref{j_xy}) take the form
\begin{eqnarray}
\label{j_parallel}
j_x &=& I B_y \left(S_1  + S_2 \frac{1+\cos{4\varphi}}{2}  \right)\:,\\
j_y &=& I B_y \left(S_2 \frac{\sin 4\varphi}{2} + \mu  \sin
2\varphi \right) \:.  \nonumber
\end{eqnarray}
These equations show that the  contribution to the
magnetic-field induced photocurrent is expected from the last term
in the right hand side of the second equation described by the
coefficient $\mu$. Fit of our data by Eqs.~(\ref{j_parallel}) presented in Fig.~3
demonstrates a good agreement between the experiment and the theory.
Furthermore, as an essential result, it follows from Fig.~3 that
the current proportional to the radiation helicity and
described by the coefficient $\mu$ is clearly
measurable. At  excitation by circularly polarized radiation the
term containing $S_2$ vanishes and the current which remains is
that due to the  term with $\mu$ only. It changes it sign by
switching from $\sigma_+$ to $\sigma_-$ as observed in
experiment.

 Microscopically the magnetic field
induced photocurrent proportional to the radiation helicity
$P_{circ}$ is caused by the spin-galvanic effect previously
reported for GaAs- and InAs-based two-dimensional structures~\cite{11}.
The effect is due to the optical orientation of carriers
followed by the Larmor precession of the oriented electronic spins
in the magnetic field and asymmetric spin relaxation processes
(for details see Refs.~\cite{6,10,11}).
Though, in general, the spin-galvanic current does not require the
application of a magnetic field, it may  be considered as a
magneto-photogalvanic effect under the above experimental
conditions.
The photocurrent described by the
coefficient $S_2 $, as well as by $S_1 $, is caused by the
magneto-gyrotropic photogalvanic effect~\cite{7,8,9}.
The microscopic mechanism of the MPGE in low-dimensional
structures has been developed most recently to describe this
effect in GaAs-, InAs- and SiGe-based structures~\cite{8,9}. It has been
shown that free carrier absorption of THz radiation results in a
pure spin current and corresponding spin separation achieved by
spin-dependent scattering of electrons in gyrotropic media. The
pure spin current in these experiments was converted into an
electric current by application of a magnetic field which
polarizes spins due to the Zeeman effect. The key experiment
supporting this microscopic mechanism is investigation of the
temperature dependence of the photocurrent: the photocurrent due
to zero-bias spin separation should be constant at low
temperatures and should behave as the ratio $n_s(T) /k_B T$ at high
temperatures~\cite{8,9}. Figure~2 shows that the temperature dependences
of $J_1 $ and $J_2 $ contributions indeed can be well fitted by
constant factors in the low temperature range and vary as $n_s(T)
/k_B T$ at high temperatures. Furthermore in the range 50-100~K
where the mobility  is almost constant the magnitude of $J_1$ changes rapidly
showing that $J$ and $\mu$ are not correlated.
The temperature behaviour of the photocurrent demonstrates the applicability of
the   discussed above model to the MPGE photocurrent observed in GaN/AlGaN
heterojunctions. We note, however, that the influence of the
magnetic field on electron scattering~\cite{Kibis1998,Kibis2000}
may also result in a
photocurrent yielding an additional contribution to the
MPGE~\cite{12}. Since the microscopic origin of both
contributions is different the relative role of them can be
clarified by additional experiments, e.g. by the variation of
$g$-factor in Mn doped nitride low-dimensional structures.

\section{Conclusion}

In summary, we  demonstrated that the presence of a
substantial gyrotropy in GaN/AlGaN heterostructures gives a root to the
spin-galvanic effect and the magneto-gyrotropic photogalvanic
effect. We note that in our GaN heterojunctions the MPGE and
spin-galvanic current contributions have the same order of magnitude.
Both effects have been proved to be an effective tool for
investigation of nonequilibrium processes, in-plane symmetry and
inversion asymmetry of heterostructures, electron momentum and spin
relaxation etc. and, therefore, may be used for investigation of
this novel material attractive for spintronics.

\section*{Acknowledgments} The financial support of the DFG, RFBR
and RSSF is gratefully acknowledged. E.L.I. thanks DFG for the
Merkator professorship.

%\section{References}

\end{document}